\documentclass[11pt]{article}

\usepackage{times}
\usepackage{bm}
\usepackage{natbib}
\usepackage{epsfig}
\usepackage{caption}
\usepackage{amssymb}
\def\pr{\mathop{\rm pr}\nolimits}	
\def\per{\mathop{\rm per}\nolimits}	
\def\cyp{\mathop{\rm cyp}\nolimits}	
\def\subdot{{\hbox{\bf .}}}

\newtheorem{theorem}{Theorem}[section]

\newtheorem{example}{Example}[section]

\begin{document}




\title{Classification based on a permanental process with cyclic approximation}

\author{Jie Yang$^{1}$, Klaus Miescke$^{1}$, and Peter McCullagh$^{2}$\\
$^1$University of Illinois at Chicago and $^2$University of Chicago}
\date{July 4, 2012}

\maketitle

\begin{abstract}
We introduce a doubly stochastic marked point process model for supervised classification problems. Regardless of the number of classes or the dimension of the feature space, the model requires only 2--3 parameters for the covariance function. The classification criterion involves a permanental ratio for which an approximation using a polynomial-time cyclic expansion is proposed. The approximation is effective even if the feature region occupied by one class is a patchwork interlaced with regions occupied by other classes. An application to DNA microarray analysis indicates that the cyclic approximation is effective even for high-dimensional data. It can employ feature variables in an efficient way to reduce the prediction error significantly.	This is critical when the true classification relies on non-reducible high-dimensional features.
\end{abstract}


\section{Introduction}

In a typical supervised or unsupervised classification problem, each observation can be treated as a single point in the feature space ${\cal X}$.  The data set is a finite point configuration $x=\{x_1, \ldots, x_n\}$ with or without class labels $y=\{y_1, \ldots, y_n\}$. A Cox process, or a doubly stochastic Poisson process (\citeauthor{cox1980} 1980; \citeauthor{kingman1993} 1993; \citeauthor{daley2003} 2003), provides a rich family of spatial point processes for aggregated point patterns. Unfortunately, for most Cox processes considered in the literature, no closed form for the distribution of $x$ is available. Markov chain Monte Carlo methods are commonly used for computational purposes. \citeauthor{pmcc2006a} (2006) introduced a special class of Cox process, the permanental process, which is fairly flexible and has a closed form for the marginal density of $x$.

\citeauthor{pmcc2006b} (2006) proposed a classification model based on the permanental process. Regardless of the number of classes or the dimension of the feature variables, the model requires only 2--3 parameters for fitting the covariance function of the random intensity. The method is effective even when the region predominantly occupied by one class is a patchwork interlaced with regions occupied predominantly by other classes. One problem of the permanental model is that it requires the calculation of ratios of weighted permanents, which is an NP-hard problem (\citeauthor{valiant1979} 1979).

In the computer science literature, the best approximation algorithm proposed by \citeauthor{bezakova2006} (2008) runs at an unappealing rate of $O(n^7\log^4n)$.  \citeauthor{kou2009} (2009) use an importance-sampling estimator to approximate weighted permanents up to a few hundred points.

We propose a different way to solve the problem. It involves a series of approximations for the weighted permanental ratio based on its cyclic expansion. The classification based on cyclic approximations works reasonably well for the examples studied.

\section{Classification Model Based on Permanental Process}
\label{classificationmodel}

\subsection{Permanental process}
\label{subsectionpermanental}

Following~\citeauthor{pmcc2006a} (2006), the permanental process on the feature space ${\cal X}$ is a Cox process with random intensity function
$$\Lambda(x_i)=\sum_{r=1}^{2\alpha} Z_r^2(x_i)\ ,$$
where $Z_1,\ldots, Z_{2\alpha}$ are independent and identically distributed Gaussian random fields with mean zero and covariance function $C/2$. For many applications, ${\cal X} = {\cal R}^d$ or ${\cal X} \subset {\cal R}^d$.

Typically, a spatial pattern consisting of $n$ points $\{x_1,\ldots, x_n\}$ is observed within a compact subset $S$, or a bounded window, in ${\cal X}$.  If $C$ is continuous on $S\times S$, it has the spectral representation
$$C(x_i,x_j)=\sum_{r=0}^{\infty} \lambda_r e_r(x_i) e_r(x_j),$$
where $\lambda_r$ and $e_r$ are the eigenvalues and the normalized eigenfunctions of $C$ on $S$, respectively. Define a new covariance function on $S$ by
$$K(x_i,x_j)=\sum_{r=0}^{\infty} \frac{\lambda_r}{1+\lambda_r} e_r(x_i)e_r(x_j).$$
We call $K$ the covariance function of the permanental process on $S\times S$. Note that $K\cong C$ if all eigenvalues are close to $0$.

The marginal density (\citeauthor{pmcc2006a} 2006, Section~3.2) of the permanental process with respect to Lebesgue measure at $x=\{x_1, \ldots, x_n\}$ is
\begin{equation}\label{marginal}
f(x)=e^{-\alpha D}\per_{\alpha}\left\{K(x)\right\}
\end{equation}
where $D=\sum_{r=0}^{\infty} \log(1+\lambda_r)$, and
\begin{equation}\label{weighted}
\per_{\alpha}\left\{K(x)\right\}=\sum_{\sigma}\alpha^{\#\sigma} K\left(x_1,x_{\sigma(1)}\right)\cdots K\left(x_n,x_{\sigma(n)}\right)
\end{equation}
is the $\alpha$-permanent of the $n\times n$ matrix $K(x)$ with components $K(x_i, x_j)$ (\citeauthor{vj1988} 1988). Here the sum runs over all permutations of $(1,\ldots,n)$ and $\#\sigma$ indicates the number of cycles. The usual permanent (\citeauthor{minc1978} 1978) corresponds to $\alpha=1$, and $\per_{-1}(A)=(-1)^n\det(A)$.

For general positive definite $K$, the permanental process is defined only for positive integer values of $2\alpha$ (\citeauthor{branden2012}, 2012), but if $K(x_i, x_j)$ is everywhere non-negative, the process can be extended to positive~$\alpha$.

Unlike general Cox processes, the permanental process has its density function in explicit form~(\ref{marginal}). The flexibility in choosing $\alpha$ and $K$ makes the permanental process potentially useful for applied work.

\subsection{Classification model with finitely many classes}
\label{finiteclassificationmodel}

For supervised classification problem with finitely many classes, the observations $x_1,$ $\ldots,$ $x_n$ come from $k$ possible classes. Assume that the observations in class $r$ follow a permanental process with parameter $\alpha_r$ and covariance function $K$ as in (\ref{marginal}).  The superposition of $k$ independent permanental processes with same $K$ is a permanental process with parameter $\alpha_\subdot=\sum_{r=1}^k \alpha_r$ and the same covariance function $K$.

\citeauthor{pmcc2006b} (2006) show that the conditional distribution of the label vector $y$ given the feature observations $x$ is
\begin{equation}
\label{ygivenx}
\pr(y\mid x)
=\frac{\per_{\alpha_1}\left\{K(x^{(1)})\right\}\cdots \per_{\alpha_k}\left\{K(x^{(k)})\right\}}{\per_{\alpha_\subdot}\left\{K\left(x\right)\right\}}\>,
\end{equation}
where $x^{(r)}$ denotes the observations belonging to class $r$ and $\per_{\alpha}\left\{K(x)\right\}$ is defined in (\ref{weighted}). Note that $\per_{\alpha}\left\{K(\emptyset)\right\}=1$ for the empty set $\emptyset$.

For a supervised classification with known label vector~$y$,	 the goal is to classify a new unit $u'$ with observed feature vector $x'$ into one of the $k$ classes. Since the conditional distribution (\ref{ygivenx}) applies to the extended sample,	the conditional distribution is given by the theorem as follows.

\begin{theorem}\label{finiteclasstheorem}
Given $x$ and $y$, the conditional probability that a new unit $u'$ with observed feature $x'$ belongs to class $r$ is
\begin{equation}\label{perratio}
\pr(u'\mapsto r \mid  x', x,y)
\>\propto\>
\frac{\per_{\alpha_r}\left\{K\left(x^{(r)}\cup x'\right)\right\}}{\per_{\alpha_r}\left\{K\left(x^{(r)}\right)\right\}}\ .
\end{equation}
If $x^{(r)}=\emptyset$, that is, no observation from class~$r$ has yet been observed, then the probability is proportional to $\alpha_r K(x',x')$.
\end{theorem}

\subsection{Classification model with infinitely many classes}
\label{infiniteclassificationmodel}

For many classification applications, for example, to identify species of animal or type of cancer, it is not appropriate to assume a finite number of classes in the population. We may consider the limit of (\ref{ygivenx}) as $k\rightarrow \infty$, $\alpha_r = \alpha\rightarrow 0$ for all $r$, and $\alpha_\subdot=k\alpha=\lambda >0$ is fixed. Fixing the number of observations $n$, the limit distribution for the unlabelled partition $B$ of $\{1,\ldots,n\}$ is
\begin{equation}
\label{Bgivenx}
\pr(B\mid \ x;\lambda)
=\lim_{\stackrel{k\rightarrow\infty}{\alpha\rightarrow 0^+}} \frac{k!\ \pr(y\mid x)}{(k-\#B)!}
=\frac{\lambda^{\#B}\prod_{b\in B}\cyp\left\{K\left(x^{(b)}\right)\right\}}
{\per_\lambda\left\{K\left(x\right)\right\}}\ ,
\end{equation}
where $\#B$ is the number of blocks of $B$, $x^{(b)} = \{x_i\mid i\in b\}$ is the set of observations belonging to block $b$, and
$$\cyp\left\{K\left(x\right)\right\}=\lim_{\alpha\rightarrow 0^+}\alpha^{-1}
\per_\alpha\left\{K\left(x\right)\right\}
=\sum_{\sigma:\#\sigma=1} K\left(x_1, x_{\sigma(1)}\right)
\cdots K\left(x_n,x_{\sigma(n)}\right)$$
is the sum of cyclic products. The product in (\ref{Bgivenx}) runs over all blocks of $B$. For example, $B=\left\{\{1,3\},\{2\},\{4,5\}\right\}$ is a partition of $\{1, 2, 3, 4, 5\}$, then the blocks of $B$ are $\{1,3\}$, $\{2\}$, and $\{4,5\}$, and the number of blocks $\#B=3$. By (\ref{Bgivenx}) and the properties of conditional probability, we have

\begin{theorem}\label{infiniteclasstheorem}
Suppose there are infinitely many classes. Given $B,x,\lambda$, the conditional probability of assigning a new unit $u'$ with feature $x'$ to block $b \in B$ is
\begin{equation}\label{cycratio}
\pr(u'\mapsto b\mid x, x', B, \lambda)
\>\propto\>
\frac{\cyp\left\{K(x^{(b)}\cup x')\right\}}{\cyp\left\{K\left(x^{(b)}\right)\right\}}\ .
\end{equation}
The conditional probability of assigning $u'$ to a new class $b=\emptyset$ is proportional to $\lambda K(x',x')$.
\end{theorem}

If $K$ is constant on ${\cal X}$, equation~(\ref{Bgivenx}) reduces to the Ewens sampling distribution (\citeauthor{ewens1972} 1972; \citeauthor{pitman2006} 2006), and expression~(\ref{cycratio}) reduces to the seating plan of a Chinese restaurant process (\citeauthor{aldous1985} 1985; \citeauthor{pitman2006} 2006).

\section{Cyclic Approximations for Permanental Ratio}
\label{cyclicappoximation}

\subsection{Approximations based on cyclic expansion}\label{sectionexpansion}

To apply the permanental classification model, we need to calculate the ratio
\begin{equation}\label{perratioR}
R_n(t;x)=\frac{\per_\alpha\{K(x\cup t)\}}{\per_\alpha\{K(x)\}}\ ,\quad \alpha>0,
\end{equation}
or to calculate the cyclic ratio
\begin{equation}\label{cycratioC}
C_n(t;x)=\frac{\cyp\{K(x\cup t)\}}{\cyp\{K(x)\}}
=\lim_{\alpha\rightarrow 0^+} R_n(t;x)
\end{equation}
for each labelled class or unlabelled block. An efficient algorithm is critical. We propose analytic approximations to the permanental ratio for classification applications.

The $\alpha$-permanent of the matrix $K[\{t, x_1, \ldots, x_n\}]$ is a sum over $(n+1)!$ terms. In a subset consisting of $n!$ terms, the index~$t$ occurs in a cycle of length~1, giving rise to the partial sum
$$
\alpha K(t, t) \per_\alpha\{K(x)\}\>.
$$
The index~$t$ may also occur in a cycle of length~2 such as $(t, x_1)$ or $(t, x_2)$ and so on. There are $n!$ permutations in which $t$ occurs in a 2-cycle, giving rise to the additional sum
$$
\sum_{i=1}^n \alpha K(t, x_i)K(x_i, t) \per_\alpha\{K(x_{-i})\}\>,
$$
where $x_{-i}$ is the set of $n-1$ points with the $i$th element removed. Similarly, the index~$t$ may occur in a 3-cycle such as $(t, x_i, x_j)$ or $(t, x_j, x_i)$, giving rise to the sum
$$
\sum_{i\neq j} \alpha K(t, x_i) K(x_i, x_j) K(x_j, t) \per_\alpha\{K(x_{-i-j})\}\>.
$$
In the cyclic expansion of the permanent of order~$n+1$, there are $n!$~terms in which $t$ occurs in a 1-cycle, $n!$~terms in which $t$ occurs in a 2-cycle, $n!$~terms in which $t$ occurs in a 3-cycle, and so on up to cycles of length~$n+1$. Therefore, we obtain the following finite expansion by cycles for (\ref{perratioR})
\begin{eqnarray*}
R_n(t; x)
&=& \alpha K(t, t) +
        \alpha\sum_i {1 \over R_{n-1}(x_i; x_{-i})} \biggl( |K(t, x_i)|^2 \\
         & &+
        \sum_{j\neq i} {1 \over R_{n-2}(x_j; x_{-i-j})} \biggl[ K(t, x_i)K(x_i, x_j)K(x_j, t)\\
         & &+
        \sum_{k\neq i,j} {1\over R_{n-3}(x_k; x_{-i-j-k})}
        \biggl\{K(t, x_i)K(x_i, x_j)K(x_j, x_k)K(x_k, t) + \cdots
\biggr\}\biggr]\biggr).
\nonumber
\end{eqnarray*}
This cyclic expansion suggests a recursive approximation in which
$$
R_n^{(0)}(t; x) = \alpha K(t, t)
$$
is the uni-cycle approximation for $n\geq 0$;
\begin{eqnarray*}
R_n^{(1)}(t; x) &=&
        \alpha K(t, t) + \alpha \sum_i {|K(t, x_i)|^2 \big/ R_{n-1}^{(0)}(x_i; x_{-i})} \\
        &=& \alpha K(t, t) + \sum_i |K(t, x_i)|^2 \big/ K(x_i, x_i)
\end{eqnarray*}
is the two-cycle approximation for $n\geq 1$;
\begin{eqnarray*}
R_n^{(2)}(t; x)
& = &\alpha K(t, t)
        + \alpha\sum_i {1\over R_{n-1}^{(1)}(x_i; x_{-i})}\biggl\{ |K(t, x_i)|^2 +
     \sum_{j\neq i} {K(t, x_i)K(x_i, x_j)K(x_j, t) \over R_{n-2}^{(0)}(x_j;
x_{-i-j})} \biggr\}\\
& = &\alpha K(t, t)
        + \alpha\sum_i \frac{|K(t, x_i)|^2}{R_{n-1}^{(1)}(x_i; x_{-i})}+
     \sum_i \frac{1}{R_{n-1}^{(1)}(x_i; x_{-i})} \sum_{j\neq i} {K(t, x_i)K(x_i, x_j)K(x_j, t) \over K(x_j,x_j)}
\end{eqnarray*}
is the three-cycle approximation for $n\geq 2$, and so on. The four-cycle approximation $R_n^{(3)}(t; x)$ for $n\geq 3$ is
\begin{eqnarray*}
 && \alpha K(t, t) +
        \alpha \sum_i {1 \over R_{n-1}^{(2)}(x_i; x_{-i})} \biggl[ |K(t, x_i)|^2 +
        \sum_{j\neq i} {1 \over R_{n-2}^{(1)}(x_j; x_{-i-j})} \times {} \\
\nonumber
                &&\biggl\{ K(t, x_i)K(x_i, x_j)K(x_j, t) + \sum_{k\neq i,j}
        {K(t, x_i)K(x_i, x_j)K(x_j, x_k)K(x_k, t) \over R_{n-3}^{(0)}(x_k;
x_{-i-j-k})} \biggr\}\biggr].
\nonumber
\end{eqnarray*}
It is natural to let
$
C_n^{(k)}(t;x)=\lim_{\alpha\rightarrow 0^+} R_n^{(k)}(t;x)
$
be the $(k+1)$-cycle approximation for $C_n(t;x)$. The two-cycle approximation $R_n^{(1)}(t; x)$ or $C_n^{(1)}(t; x)$ is a kernel function, which is an additive function of $x$, while the three-cycle approximation is not.

\bigskip\noindent
For $n=0$ or $x=\emptyset$, $R_n^{(0)}(t;x)=\alpha K(t,t)=R_n(t;x)$ is exact. For $n=1$,
\[
R_n^{(1)}(t;x)
=  \alpha K(t, t) + \alpha \sum_i \frac{|K(t, x_i)|^2}{R_0^{(0)}(x_i; x_{-i})}
=  \alpha K(t, t) + \alpha \sum_i \frac{|K(t, x_i)|^2}{R_0(x_i; x_{-i})}
= R_n(t; x).
\]
In both cases, $C^{(n)}_n(t;x)\equiv C_n(t;x)$. By induction, we obtain in general

\begin{theorem}\label{rnnexacttheorem}
For $n=0,1,2,\ldots$,
$R^{(n)}_n(t;x)\equiv R_n(t;x)$, and
$C^{(n)}_n(t;x)\equiv C_n(t;x)$.
\end{theorem}

\noindent
Up to $k=3$, that is, the four-cycle approximation, $R_n^{(k)}(t;x)$ is easy to compute, even for fairly large values of~$n$. The time complexity is $O(n)$ for the two-cycle approximation, $O(n^2)$ for the three-cycle approximation, and $O(n^3)$ for the four-cycle approximation. For some special cases, the cyclic approximation provides an exact value for $R_n(t;x)$.

\begin{example}\label{example31}
Let $K(t,t')=\delta_{tt'}f(t)$, which corresponds to diagonal matrices. Here $f$ is some positive non-random function on ${\cal X}$, and $\delta_{tt'}=1$ if $t=t'$ and $0$ otherwise. If $t, x_1, \ldots, x_n$ are pairwise different, then for each $k=0,\ldots, n$,
$$R_n(t;x)=R_n^{(k)}(t;x)\equiv \alpha f(t), \>\>\>C_n(t;x)=C_n^{(k)}(t;x)\equiv 0.$$
\end{example}

\begin{example}\label{example32}
Let $K(t,t')\equiv c$ for some constant $c>0$, which corresponds to constant matrices. Then $\per_\alpha\{K(x)\} = c^n\alpha (\alpha+1) \cdots (\alpha+n-1)$. For each $k=1,\ldots,n$,
$$R_n(t;x)=R_n^{(k)}(t;x)\equiv c(\alpha+n), \>\>\>C_n(t;x)=C_n^{(k)}(t;x)\equiv cn.$$
Note that $R_n^{(0)}(t;x)\equiv c\alpha$, $C_n^{(0)}(t;x)\equiv 0$.
\end{example}

\begin{example}\label{example33}
Let $K$ be a projection of rank~$\nu$ on ${\cal X}$.  That is,
\[
\int_{\cal X} K(t,t)\mu(dt)=\nu,\qquad
\int_{\cal X} K(s,t)K(t,u)\mu(dt)=K(s,u).
\]
Then the two-cycle approximation determines a probability density in the sense that it is non-negative and has unit integral:
\begin{eqnarray*}
(n+\alpha \nu)^{-1} \int_{\cal X} R_n^{(1)}(t; x)\, \mu(dt) &=&
        (n+\alpha\nu)^{-1} \biggl\{\alpha\nu + \sum_i \int \frac{|K(t, x_i)|^2}{K(x_i, x_i)} \,
        \mu(dt)
\biggr\} \\
        &=& (n+\alpha\nu)^{-1} \biggl\{\alpha\nu + \sum_i \frac{K(x_i, x_i)}{K(x_i, x_i)} \biggr\} \\
        &=& 1.
\end{eqnarray*}
A similar argument shows that the three-cycle and four-cycle approximations also integrate to unity, but it is not clear whether they are non-negative.
\end{example}

\begin{theorem}\label{exactratiotheorem}
Suppose $n\geq 2$.
(i) If the $n\times n$ matrix $K(x)$ is diagonal, then
$$R_n(t;x)=R_n^{(1)}(t;x)=\cdots=R_n^{(n)}(t;x)=\alpha K(t,t)
+ \sum_{i=1}^n\frac{|K(t,x_i)|^2}{K(x_i,x_i)}\ .$$
(ii) If $K(x_i,x_j)\equiv c$, $i,j=1,\ldots,n$, $c \neq 0$, then for $k=2, \ldots, n$,
\[
R_n(t;x) = R_n^{(k)}(t;x)=\alpha K(t,t)+
\alpha\sum_{i=1}^n\frac{|K(t,x_i)|^2}{c(\alpha+n-1)}
+\sum_{\stackrel{i,j=1}{i\neq j}}^{n}\frac{K(t,x_i)K(t,x_j)}{c(\alpha+n-1)}\ .
\]
(iii) Suppose $K(x)$ is block-diagonal with constant blocks. That is, there exist a partition $B$ of $\{1,2,\ldots, n\}$ and some constants $c_b\neq 0, b\in B$, such that, $K(x_i,x_j) = c_b$ if $i,j\in b$, and $0$ otherwise. Then for $k=2, \ldots, n$,
\[
R_n(t;x) = R_n^{(k)}(t;x)=\alpha K(t,t)+
\alpha\sum_{b\in B}\frac{\sum_{i\in b}|K(t,x_i)|^2}{c_b(\alpha+|b|-1)}
+\sum_{\stackrel{b\in B}{|b|\geq 2}}\frac{\sum_{\stackrel{i,j\in b}{i\neq j}}K(t,x_i)K(t,x_j)}{c_b(\alpha+|b|-1)}\ .
\]
\end{theorem}

\noindent
Based on Theorem~\ref{exactratiotheorem}, the three-cycle or higher order cyclic approximation is exact if the $n\times n$ matrix $K(x)$ is diagonal, constant, or block-diagonal with constant blocks. The $(n+1)\times(n+1)$ matrix $K(t\cup x)$ may not be diagonal, constant, or block-diagonal.

\subsection{Accuracy of the cyclic approximations}

For $n < 20$, the accuracy of the approximation can be checked directly by comparison with the exact computation. Our experience is that the three-cycle approximation is adequate in this range, and the four-cycle approximation usually has negligible error. For larger values, say $n > 50$, the accuracy can be checked by examining special cases in which the permanent can be calculated exactly in reasonable time. For example, to calculate the $\alpha$-permanent of a penta-diagonal matrice $A$, that is, $A_{i,j}=0$ for $|i-j|>2$, three-cycle or higher order cyclic approximation is essentially exact. For more general matrices, the accuracy can be gauged to some extent from an examination of the sequence of approximations.

\begin{figure}\label{fig1}
\begin{center}
\psfig{figure=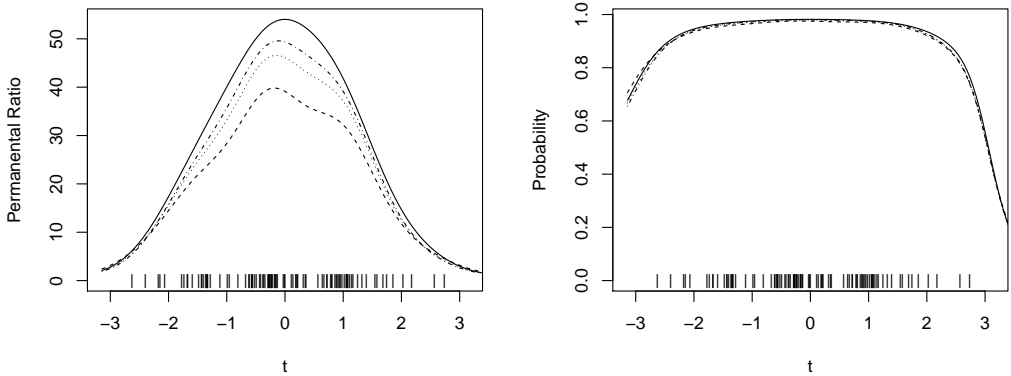,height=2.5in,width=5.5in,angle=0}
\vskip -0.3cm
\captionsetup{width=.75\textwidth}
\caption{Approximations of the permanental ratio $R_n(t; x)$ (left panel) and exact probability that $t$ belongs to class 1 (right panel) from Kou and McCullagh's estimate (solid), four-cycle (dot-dash), three-cycle (dot), and two-cycle (dash) approximations}
\end{center}
\end{figure}

The left panel of Figure~1 shows the approximate values of the permanental ratio (\ref{perratioR}) for a sample of 100 $x$-values in $(-\pi, \pi)$, plotted as a function of $t$ in the same range. The 100 points are generated from the symmetric triangular distribution on $(-\pi, \pi)$. For this example, $\alpha = 1$, and $K(t, t') = \exp\{-(t - t')^2/\tau^2\}$ with $\tau=1$. In the central peak, the lowest curve is the two-cycle approximation, and the next two curves are successive approximations up to four-cycle. The highest curve is the estimated values from the importance sampler described by \citeauthor{kou2009} (2009, Section~4). The shape of these relative intensity functions depends fairly strongly on the value of $\tau$, but only slightly on $\alpha$. In all cases, the difference between the three-cycle and four-cycle approximations is considerably smaller than the difference between the two-cycle and three-cycle ones. For $\alpha=\tau=1$, the four-cycle approximation is approximately 6\% larger than the three-cycle in the central peak, while the three-cycle approximation is approximately 18\% larger than the two-cycle one. On average, the relative differences between the cyclic approximations and \citeauthor{kou2009}'s importance sampling estimate are 19\% for two-cycle, 12\% for three-cycle, and 10\% for four-cycle approximations, respectively.

To check the performance of our cyclic approximations for supervised classification applications, we generate another 100 points from the symmetric triangular distribution on $(\pi, 3\pi)$ denoted by class~2 and regard the first 100 points shown in Figure~1 as class~1's. According to expression~(\ref{perratio}), we can calculate the probability that a point with feature $t$ belongs to class~1. The right panel of Figure~1 plots the probabilities when the permanental ratios are calculated based on the cyclic approximations or \citeauthor{kou2009}'s importance sampler. The differences among the four approximations are negligible. The maximum relative differences between the cyclic approximations and \citeauthor{kou2009}'s estimate are 4.3\% for two-cycle, 3.4\% for three-cycle, and 3.3\% for four-cycle, respectively. If we regenerate class~2 from a symmetric triangular distribution on $(0.5\pi, 2.5\pi)$ which is overlapped with the region of class~1, the maximum relative differences can be as large as $44\%$ and $14\%$ for the two-cycle and three-cycle approximations, while the four-cycle approximation still works reasonably well with a maximum relative difference $4.2\%$. The worst cases usually occur at the boundary or the overlapped part $(0.5\pi, \pi)$. Even for the overlapped distributions, the corresponding maximum absolute differences between the cyclic approximations and \citeauthor{kou2009}'s estimate are $0.045$ for two-cycle, $0.018$ for three-cycle, and $0.023$ for four-cycle approximations in terms of class probability.

As for computation time, it took a personal computer with 2.8GHz CPU and 2GB RAM 1.3 seconds in total to finish all calculations based on two-cycle, three-cycle and four-cycle approximations, or about 700 seconds based on \citeauthor{kou2009}'s importance sampler with sample size 20,000.

\section{A Simulated Example}
\label{simulatedexample}

We use an artificial example to illustrate how the proposed model with cyclic approximation works for a supervised classification problem. This example has two classes in a $3$ by $3$ chequer-board layout with classes labelled as follows.
\begin{center}
\begin{tabular}{|c|c|c|}\hline
1 & 2 & 1\\ \hline
2 & 1 & 2\\ \hline
1 & 2 & 1\\ \hline
\end{tabular}
\end{center}
The training dataset consists of 90 units, with 10 feature values uniformly distributed in each $1$~by~$1$ small square, as shown in Figure~2. We assume the two-class model based on permanent processes with $\alpha_1=\alpha_2=\alpha$ and covariance function $K_1(t, t') = \exp(-\|t - t'\|/\tau)$ or $K_2(t, t') = \exp(-\|t - t'\|^2/\tau^2)$. The calculations are based on the four-cycle approximation for the permanental ratio described in Section~\ref{sectionexpansion}. The parameters $\alpha$ and $\tau$ are chosen by $10$-fold cross-validation.

\begin{figure}\label{fig2}
\begin{center}
\psfig{figure=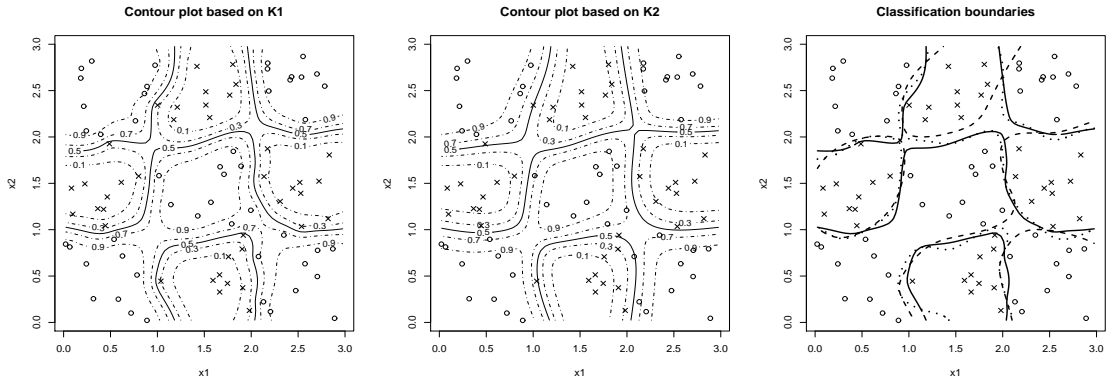,height=5.9in,width=2.12in,angle=270}
\vskip -0.3cm
\captionsetup{width=.80\textwidth}
\caption{Classification results. The contour plots of the probability that a new point is assigned to class~1 (round dots) based on permanental models $K_1$ and $K_2$ are shown in the left and middle panels. The boundary lines of classification based on $K_1$-permanental model (solid), neural network (dash), or support vector machine (dot) are shown in the right panel.}
\end{center}
\end{figure}

The left and middle panels of Figure~2 provide the contour plots of the probability that a new point is assigned to class~1. For the parameter values chosen, the range of predictive probabilities depends, to a moderate extent, on the configuration of $x$-values in the training sample, but the extremes are seldom below 0.1 or above 0.9 for a configuration of 90 points with 10 in each small square. The range of predictive probabilities decreases as $\alpha$ increases, but the 50\% contour line, that is, the solid line in Figure~2, is little affected, so the classification is fairly stable. The class boundaries based on $K_1$ and $K_2$ are slightly different. The boundary based on $K_1$ is more sensitive to the boundary points.  In practice, one may use cross-validation to choose the optimal type of covariance function from several candidates. In this case, $K_1$ works slightly better according to error rate and cross-entropy loss.

The right panel of Figure~2 compares the class boundaries generated by the proposed model with $K_1$, a neural network method using single layer with $18$ hidden units and weight decay $0.001$ chosen by cross-validation, and a support vector machine using Gaussian kernel with tuning parameter chosen by cross-validation. Since we know the data-generating mechanism, we can evaluate the performance, and the $K_1$-permanental model performs best.

\begin{center}
Table 1: Error counts out of $90$ and $3600$ respectively\\
\begin{tabular}{l|c|c}\hline
Classifier & Training error & Testing error\\ \hline
Proposed model with $K_1$ & 0 & 308\\
Proposed model with $K_2$ & 5 & 301\\
Neural network & 0 & 334\\
Support vector machine & 0 & 357\\
Aggregate classification tree & 0 & 391\\
$k$-nearest neighbor & 6 & 412\\ \hline
\end{tabular}
\end{center}

Given that the correct classification is determined by the chequerboard rule, the error rates for training data and $60\times 60$ grid points serving as testing data are summarized in Table~1.  For comparison purposes, some commonly used classifiers are listed in Table~1 too. In addition to the neural network method and support vector machine, we also check the results based on an aggregated classification tree with bagging number 100 and a $k$-nearest neighbor classifier with $k=5$ chosen by cross-validation. Diagonal linear discriminant analysis and logistic regression do not work for the original $x_1$ and $x_2$ in this case, because the class regions are non-convex and interlaced.

\section{Microarray Analysis: Leukemia Dataset}
\label{microarray}

The leukemia dataset described by \cite{golub1999} uses microarray gene expression levels for cancer classification. It consists of 72 tissue samples from two types of acute leukemia, 47 samples of type ALL and 25 of type~AML. The version used here, from the  {\tt R} package {\tt golubEsets} downloaded from {\tt http://bioconductor.org}, contains expression levels for 7129 genes in each of 72 tissue samples.

\begin{figure}\label{fig4}
\begin{center}
\psfig{figure=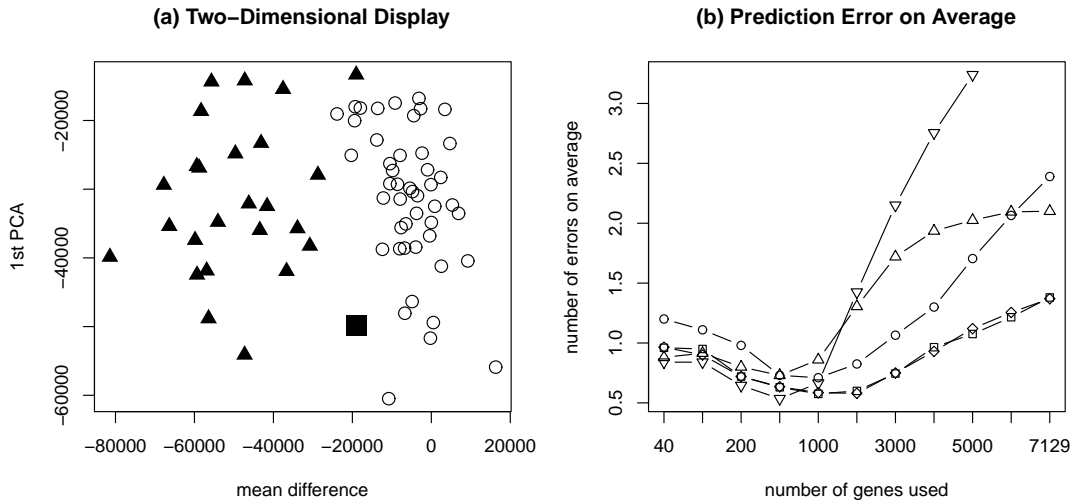,height=2.9in,width=5.8in,angle=0}
\vskip -0.3cm
\captionsetup{width=.80\textwidth}
\caption{Leukemia Data. Left panel: a two-dimensional display of the dataset with 47 ALL (round dot), 25 AML (triangle), and a new observation (square). Right panel: number of test errors on average over 200 learning/testing partitions based on different methods, including support vector machine (point-down triangle), diagonal linear discriminant analysis (point-up triangle), $k$-nearest neighbor (round dot), permamental model with $K_2$ covariance (diamond), permanental model with $K_1$ covariance (square, overlapped with diamond). }
\end{center}
\end{figure}

The left panel of Figure~3 shows a two-dimensional projection in which the $x$~axis is the straight line joining the class centroids, and the $y$~axis is the first principal component. Unlike the usual heatmap display such as Fig.~3B in \cite{golub1999}, each sample is plotted here as a single point. The goal is to classify each new tissue sample as ALL or AML based on the gene expression levels.

The leukemia dataset has been widely used for testing classifiers.  \cite{dudoit2002} did a comprehensive comparison of various discriminant methods using this dataset as well as two other popular microarray datasets. Based on their study, the nearest neighbor classifier and the diagonal linear discriminant analysis work the best when $40$ selected genes are considered.

To compare the performance of the proposed method with other methods, we follow the training/testing partitioning procedure used by \cite{dudoit2002}. The 72 samples are randomly divided into 48 training points and 24 testing points.  Each classifier is fitted or trained using the 48 training points and tested using the 24 testing points.  The number of misclassified points out of 24 is recorded. The procedure is repeated 200 times for each classifier.  The number of test errors on average is used to evaluate the performance of classifiers.

The right panel of Figure~3 shows the number of prediction errors on average over 200 random training/testing partitions. The genes used for discriminant analysis are selected according to the ratio of between-group variance to within-group variance \cite[Section~3.4]{dudoit2002}.  The proposed models with $K_1$ or $K_2$ are compared with the two winners, $k$-nearest neighbor and diagonal linear discriminant analysis methods,  in \cite{dudoit2002}, as well as the support vector machine method which became popular more recently. As the number of selected genes increases, the mean number of test errors of the four classifiers follows a similar pattern.  It decreases initially as more information becomes available for parameter estimation, but subsequently increases as the signal becomes lost in the noise. The proposed models with $K_1$ and $K_2$ perform as well as the support vector machine, but better than the $k$-nearest neighbor and diagonal linear discriminant analysis methods, in the sense of minimum average error count. Compared with the support vector machine, the proposed model performs reasonably well even with bad selection of covariates. It seems more capable of handling high-dimensional data. This is critical when the true classification relies on non-reducible high dimensional features. In terms of computational time, the proposed method is comparable with the neural network and support vector machine methods with moderate data size, but slower than the diagonal linear discriminant and $k$-nearest neighbor methods. As the number of feature variables increases, the error rates increase for all classifiers, but more rapidly for the neural network and support vector machine than for either permanental classifier.

\section*{ACKNOWLEDGEMENT}
The authors thank the editor, the associate editor, and the referee for valuable comments. This research was supported by grants from the U.S. National Science Foundation.

\appendix
\section*{Appendix}

{\bf Proof of Theorem~\ref{exactratiotheorem}}
We only need to prove case~(iii). Because case~(i) corresponds to $|B|=n, |b|\equiv 1$, while case~(ii) corresponds to $|B|=1, |b|=n$.

First if $K(t \cup x)$ is also block-diagonal, then $R_n(t;x) = R_n^{(k)}(t;x) = c_b (\alpha + |b|)$
given $t \in b$, or $=\alpha K(t,t)$ given that $t$ does not belong any block of $B$. Here $k = 1,2, \ldots, n$. Therefore, $R_n(x_i;x_{-i}) = R_n^{(k)}(x_i;x_{-i}) = c_b (\alpha + |b| - 1)$ given $i \in b$; $R_n(x_j;x_{-i-j}) = R_n^{(k)}(x_j;x_{-i-j}) = c_b (\alpha + |b| - 2)$ given $i,j \in b, i\neq j$; and so on.

The formula for $R_n(t;x)$ in case~(iii) can be justified by applying mathematical induction on the cyclic expansion of $R_n$. For its cyclic approximations, $R_n^{(1)}(t;x) = \alpha K(t,t) + \sum_{b\in B} c_b^{-1}\sum_{i \in b} |K(t,x_i)|^2 \neq R_n(t;x)$. It is straightforward to verify that $R_n^{(2)}(t;x) = R_n(t;x)$. The formula for $R_n^{(k)}(t;x)$ with $k\geq 3$ can be justified using the equation below with index $l=1,2,\ldots,k-2$.
\begin{eqnarray*}
& &\hspace{-1cm}\sum_{i_{k-l}\neq i_1,\ldots,i_{k-l-1}} \frac{K(t,x_{i_1})K(x_{i_1},x_{i_2})\cdots K(x_{i_{k-l}},t)}{c_b \alpha}\\
& &=
\sum_{i_{k-l}\neq i_1,\ldots, i_{k-l-1}}\frac{1}{R_{n-k+l}^{(l)}(x_{i_{k-l}}; x_{-i_1\cdots -i_{k-l}})}
\biggl\{K(t,x_{i_1})K(x_{i_1},x_{i_2})\cdots K(x_{i_{k-l}},t) +\\
& &\quad\sum_{i_{k-l+1}\neq i_1,\ldots,i_{k-l}}\frac{1}{c_b \alpha}\
K(t,x_{i_1})K(x_{i_1},x_{i_2})\cdots K(x_{i_{k-l+1}},t)\biggr\},\quad i_1,\ldots,i_k \in b
\end{eqnarray*}
\hfill{$\Box$}

\end{document}